\documentclass[aps,prd,reprint]{revtex4-1}
\usepackage{graphicx}%
\usepackage{hyperref}
\usepackage{amssymb}
\usepackage{amsmath}

\begin{document}
\title{Matter Density versus Distance for the Neutrino Beam from Fermilab to
Lead, South Dakota, and Comparison of Oscillations with a Variable and
a Constant Density}
\author{Byron Roe} 
\affiliation {University of Michigan, Ann Arbor, MI 48109-1040}
\date{\today}
\begin{abstract}
This paper is divided into two parts.  In the first part, 
the material densities passed through for neutrinos going from FNAL
to Sanford Laboratory are calculated using two recent density tables, Crustal
[G.~Laske,  G.~Masters. Z.~Ma, and M.~Pasyanos, Update on CRUST1.0 -- A 1-degree global model of Earth's crust, Geophys. Res. Abstracts {\bf 15}, EGU2013-2658 (2013);
For the programs and tables, see the  website: 
\url{http://igppweb.ucsd.edu/~gabi/crust1.html}.]
and Shen-Ritzwoller [W.~Shen and M.~H.~Ritzwoller. Crustal and uppermost mantle structure beneath the United States, J. Geophys. Res.: Solid Earth {\bf 121}, 4306 (2016)],
as well as the values from an older table PEMC [A.~M.~Dziewonski, A.~L.~Hales and E.~R.~Lapwood, Parametrically simple earth models consistent with geophysical data, Phys. Earth Plan. Int.
{\bf 10} 12 (1975)];
For further information see the website: \url{http://ds.iris.edu/ds/products/emc-pem/}.]
 In the second part, neutrino oscillations at Sanford Laboratory are examined for the
variable density table
of Shen-Ritzwoller.
These results  are then compared with oscillation results using 
the mean density from the Shen-Ritzwoller tables and one other fixed density.
For the tests made here, the mean density results are quite similar to those found 
using the variable density vs distance.
\end{abstract}
\pacs{xxx}
\maketitle
\section{Introduction}
The Long-Baseline Neutrino Facility (LBNF) \cite{Barger:2007yw} and the Deep Underground
Neutrino Experiment (DUNE) \cite{dune},
now under
preliminary construction will send a beam of neutrinos from the Fermi National
Accelerator Laboratory (FNAL) near Chicago to the Sanford Laboratory located in a former
gold mine in Lead, South Dakota.  The neutrino beam will travel through varying densities
of material along its path.  Along its way the neutrinos will oscillate between the three
known kinds of neutrinos.  This oscillation is affected by the presence of the material
or, more precisely, by the density of electrons along its path \cite{msw}.
Although it is possible to calculate the oscillations expected on a variable density
path, most of the preliminary calculations have assumed a constant average density.
An early LBNF report \cite{Barger:2006vy} stated that to include the effects of variable density, a 5\% density systematic was assumed.

In the first part of this paper, the variable density travelled by the neutrinos along their
path is calculated using two recent density tables, Crustal\cite{crustal} 
and Shen-Ritzwoller\cite{Colorado}, 
as well as the values from an older table PEMC\cite{pemc}. The method of calculation here 
can
be used as a template for finding the densities along other long neutrino beams.

In the second part  of this paper, oscillations calculated using the variable density 
path are compared with two fixed density calculations.
\section{Finding Densities along the Neutrino Path}
\subsection{Dividing Up The Path}
The earth is approximately an ellipsoid \cite{ellipsoid}.  The radius in the polar
direction is 6356 km and  in the equatorial direction is 6378 km. Both of these numbers are
accurate to better than 0.1 km.

Twenty five points were selected taking equal intervals of latitude (lat)
and longitude.  For two points at the same latitude, the distance between the
two longitude points is not constant, but varies as $\cos(\rm lat)$ going from zero
at the poles to a maximum at the equator. For the DUNE beam path, the adjacent
points have slightly different latitudes.  However, the latitude differences between adjacent points
are quite small and taking a mean value between adjacent points introduces a negligible error.

Let the distance from the center of the earth to sea-level at a given
latitude-longitude value be $RL_i$, the local radius at point $i$. For $i>1$
let $\Delta\theta_i$ be the angle between  $RL_i$ and
$RL_{i-1}$, and $\theta_i$ be the total angle between the initial local
radius ($RL_1$) and $RL_i$.
\begin{equation}
(x/6378)^2 + (y/6356)^2 = 1.
\end{equation}

Then $x_i = RL_i\cos({\rm lat});\ y = RL_i\sin({\rm lat}).$
\begin{equation}
1/RL_i = \sqrt{(\cos({\rm lat})/6378.)^2+ (\sin({\rm lat})/6356.)^2).}
\end{equation}

If we have a flat earth then then we would go from the initial height to
final height linearly with distance (${\rm dist}(i)$) along the neutrino beam.  Let ${\rm fltosl}$ be the distance along the neutrino beam from FNAL to Sanford Laboratory.
\begin{eqnarray}
&&{ \rm flat\  height}(i)  = ({\rm endseaheight *dist}(i) +  {\rm startseaheight} \nonumber \\
&& *  {\rm (fltosl - dist}(i)))/{\rm  fltosl}.
\end{eqnarray}
The start height of the beam at FNAL is 228.4 m above 
sea level and the end height of the midpoint of the detector at Sanford Laboratory is 159 m.

For the curved earth part starting and ending at sea-level with a total arc of
$\theta_{total}$, the angle of the arc is taken from $-\theta_{total}/2$ to $+\theta_{total}/2$.
For 25 points, the midpoint in the neutrino beam path would be given by point 13 if
the latitude of FNAL and Sanford Laboratory were the same.  In fact, they are at different latitudes 
and that introduces a non-symmetric change in the path segment lengths along the
beam path, which changes the center point slightly.  Empirically it is found to be located 2\% of the way
between point 13 and point 14, ``point 13.02".

See Fig. 1. Let  $L$ be the straight line connecting
the sea level points at initial and final destinations,  $R$ be the local radius at the center of the
beam path, $s$ 
be the perpendicular distance from the midpoint of $L$ to the circle (the sagitta),
and $t$ be the distance along the local radius from a point on L at a distance $d$ from the start 
to the local circle.
\begin{align} 
R^2 = (R-s)^2 + (L/2)^2;\cr
(R -s)^2 = R^2 -(L/2)^2.
\end{align}
$t$ is not quite perpendicular to the straight line $L$, but the error is small.
The fractional error in $t$ is zero at the center of the arc and increases, approximately quadratically,
approaching a value of 0.5\% of the perpendicular distance by the end
of the arc, where $t$ is very small. 
\begin{align}
(R-s)^2 + (d-L/2)^2 = (R-t)^2.
\end{align}
Substitute Equation 4 into Equation 5. 
\begin{align}
R^2 -(L/2)^2 + (d-L/2)^2 = (R-t)^2;\cr
(L/2)^2-(d-L/2)^2  = 2Rt -t^2.
\end{align}
Ignore the $t^2$ term.
\begin{align}
t = [(L/2)^2-(d-L/2)^2]/(2R).
\end{align}
For the calculation of $t$, the variation of the local radius over the path segment from $i$ to $i+1$ is produces
a negligible effect.
The distance above sea level at distance $d$ is then given
by the sum of the flat height and the curved height ($t$).
There is an additional
effect called the geoid height \cite{geoid}, but it is very small, about 0.01 m for the FNAL 
point and $-13.7$ m for the Sanford Laboratory point.

\begin{figure}
\includegraphics[height=8.cm]{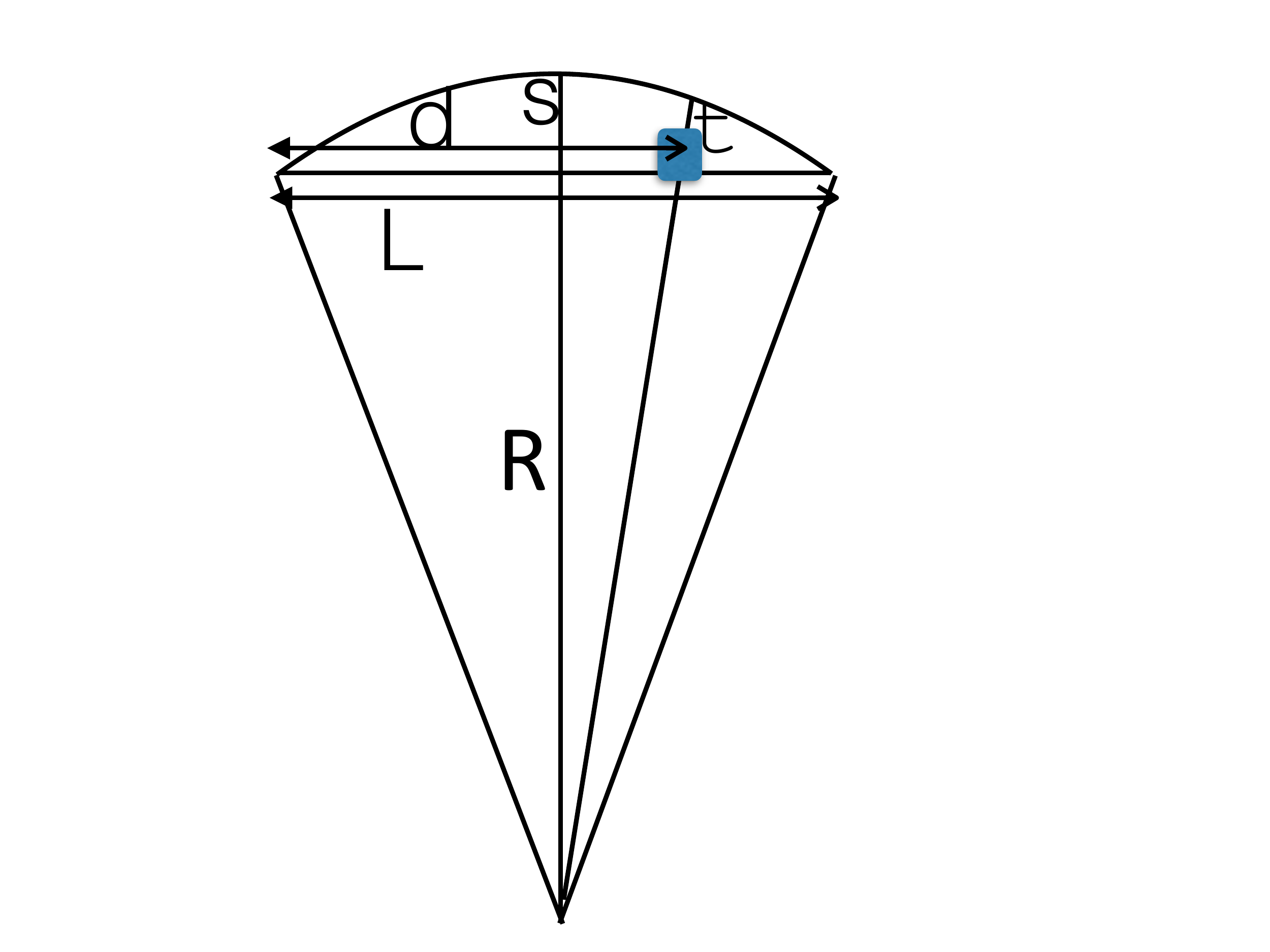}
\caption{ Figure to find the height of the earth surface above the straight line $L$ connecting
the sea level points at initial and final destinations.  $R$ is the local radius of the circle
at the center of the beam path, $s$ is 
the perpendicular distance from the midpoint of $L$ to the circle (the sagitta),
and $t$ is the distance along the local radius from a point on L a distance $d$ from the start 
to the local circle.
}
\label{Figure 1}
\end{figure}

Let  $\theta_{\rm midpoint}$ be the angle
between the local radius for point 1, and the midpoint radius.
For point $i$, the angle that $t$ makes with the midpoint radius is $\theta_i-\theta_{\rm midpoint} = \alpha$.  This angle is also the angle that the tangent to the local radius circle makes with the line $L$.
For this short segment the length of the arc and the length of the chord are essentially equal.

For $i>1$, the straight line distance from FNAL to Sanford Laboratory is incremented by 
\begin{equation}
{\rm dist}(i) = {\rm dist}(i-1) +\cos(\alpha)\times
 RL_i\times\Delta\theta_i.
\end{equation}
 The distance from FNAL to Sanford Laboratory seen by the
 neutrino beam (fltosl) is calculated to be ${\rm fltosl}=1284.9$ km.

The density maps depend on the  depth of the beam below ground at the various points.
At Sanford Laboratory there are a number of hills and the beam ends up above sea level even though the
center of the detector is
close to 1470 m beneath the surface. The elevation at a given latitude and longitude can be 
obtained from a convenient web site \cite{FreeMap} and the difference between the elevation and the sea level
height of the beam is then the depth.  See Fig. 2.  In general the elevation varies smoothly except very near to Sanford Laboratory.  If the elevation had fluctuated considerably
over a fair fraction of the path
it would have added uncertainty to the density map.
\begin{figure}[tbp] 
\includegraphics[height=7cm]{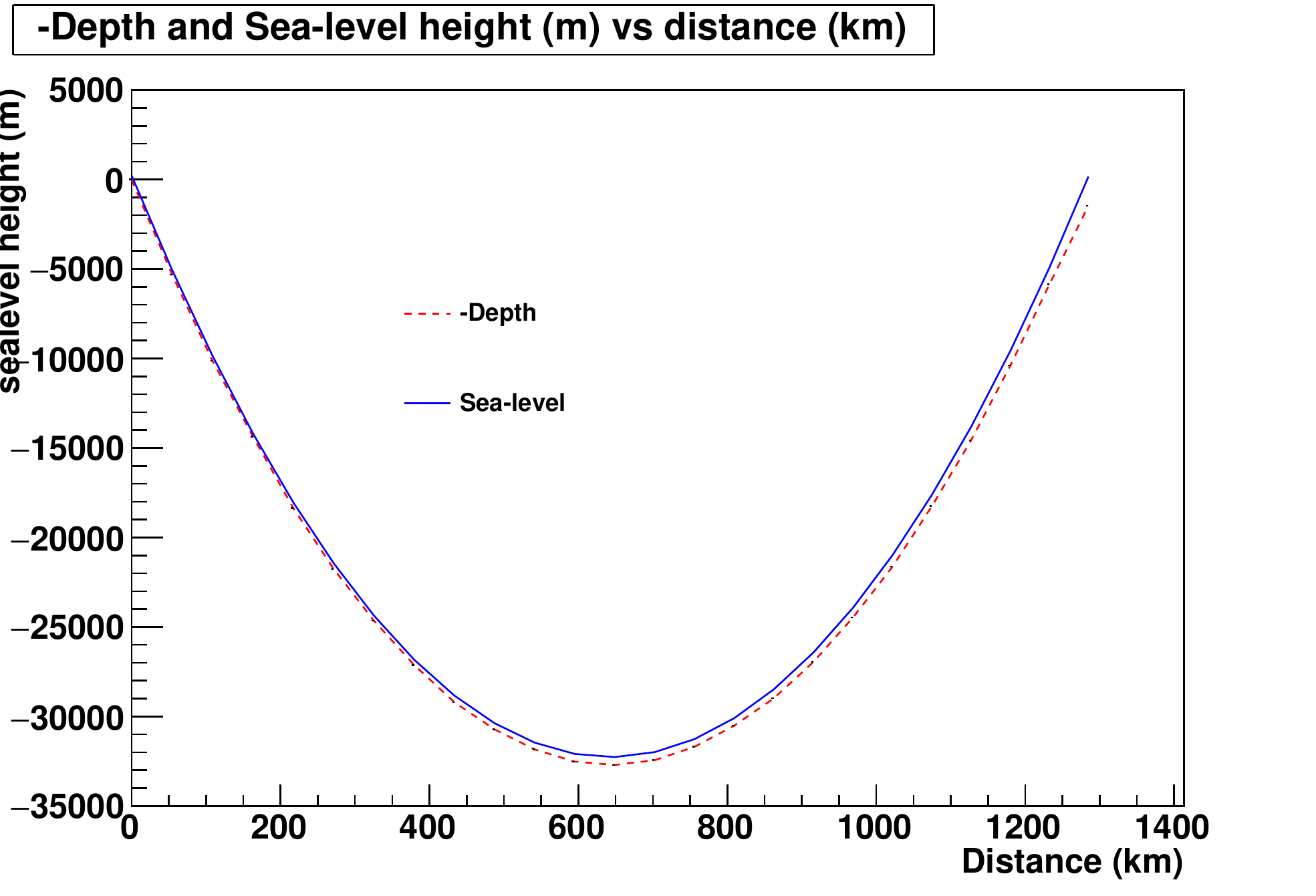}
\caption{ Sea height and negative depth vs distance from Fermilab for the neutrino beam.  
The solid line (blue) is the sea height and the dashed line (red) is the negative depth.
}
\label{Figure 2}
\end{figure}

\subsection{Results And Their Uncertainties}
Crustal is a recent (2013) attempt to find the density of the earth as a function of latitude
and longitude. CRUST1.0 is an 8 layer model. Although it is not needed here, a ninth layer gives 
the density below the Moho. 
Crustal
averages crust structure over $1\times1$ degree cells (about 110$\times$110 km). 
The map is based on the ETOP01 global relief model  produced by the National Centers for
Environmental information, a part of the National Oceanic and Atmospheric Administration\cite{NOAA}.

The
 model is defined from +89.5 to $-$89.5 deg. latitude and $-$179.5 to +179.5 deg.
longitude. 
Density is in gm/cm$^3$.  Our longitude (W) corresponds to negative values here.
Crustal supplies a program (getCN1point) which 
for a given latitude and longitude at the midpoint of a cell, gives the density of each layer and  the
bottom of the layer.
For all maps in this paper, the depth, not the sea-level height is used in the maps.

The Shen-Ritzwoller model is a new (2016) density map only of the United states in $1/4\times 1/4$ degree cells of latitude and longitude.  The density map is divided into many more
layers, than the Crustal map. There are more than 50 layers.  

There is also an older map, PEMC included for historical reasons.  A comparison of
the density vs distance results of each map is shown in Fig. 3 and the numerical results are given in
Tables I and II.


\begin{table} 
\begin{center}
\begin{tabular}   {|   l  |  l  |  l |   l | l  | l    l    |} \hline \hline
  Num. & Lat. &  Long. &   Distance &  Sea height  &    Depth & \\ \hline \hline
    1    &      41.833   &   268.272   &     0.000  &    228.444  &     --2.244 & \\
    2    &      41.938  &    268.918   &    54.379  &  --5048.751  &   5310.851 & \\ 
    3     &     42.043  &    269.563   &   108.714  &  --9852.368  &  10129.269 & \\
    4      &    42.148  &    270.209   &   163.003  & --14184.244  &  14364.145 & \\
    5    &      42.253  &    270.854  &    217.240  & --18046.264 &   18360.764 & \\
    6     &     42.359   &   271.500  &    271.421  & --21440.344  &  21756.344 & \\
    7      &    42.464    &  272.145   &   325.542  & --24368.449  &  24652.648 & \\
    8     &     42.569   &   272.791  &    379.599  & --26832.572 &   27128.373 & \\
    9     &     42.674   &   273.436   &   433.588  & --28834.752  &  29206.652 & \\
   10     &    42.779  &    274.082  &   487.504  & --30377.055  &  30720.654 & \\
   11     &     42.884   &   274.727  &    541.344  & --31461.594  &  31838.994 & \\
   12     &     42.989   &   275.373  &    595.102  & --32090.506  &  32519.906 & \\
   13     &     43.094    &  276.019   &   648.776  & --32265.973  &  32706.572 & \\
   14    &      43.200   &   276.664  &    702.362  & --31990.203  &  32440.703 & \\
   15     &     43.305   &   277.310  &    755.855 &  --31265.445 &   31693.746 & \\
   16     &     43.410  &    277.955  &    809.251 &  --30093.979  &  30513.578 & \\
   17     &     43.515   &   278.601  &    862.547 &  --28478.111 &   28977.512 & \\
   18    &      43.620   &   279.246  &    915.739 &  --26420.191 &   26946.592 & \\
   19    &      43.725  &   279.892   &   968.823 &  --23922.588 &  24466.488 & \\
   20    &      43.830  &    280.537   &  1021.795 & --20987.715  &  21628.814 & \\
   21     &     43.936   &   281.183   &  1074.652 &  --17618.004  &  18252.004 & \\
   22     &    44.041  &    281.828  &   1127.390 &  --13815.924   & 14566.324 & \\
   23      &    44.146  &    282.474   &  1180.005  &  --9583.969  &  10398.169 & \\
   24     &    44.251  &    283.119   &  1232.494  &  --4924.664  &   5860.664 & \\
   25      &    44.356  &    283.765 &    1284.852  &    159.438 &    1468.962 & \\ \hline \hline
 \end{tabular}  
 \caption{ The columns describe
point number, latitude, longitude, distance along the beam from the start at Fermilab (km),
sea level height (m) (usually negative), and depth, {\it i.e.}, the distance below earth's crust (m).
}
\label{Table 1}
\end{center}
\end{table}

\begin{table} 
\begin{center}
\begin{tabular}   {|   l  |  l  |  l |   l | l  | l    l    |} \hline \hline
   Num. & Depth  & Sea height & $\rho_{CRU}$ & $\rho_{SR}$ & $\rho_{PEMC}$  & \\ \hline \hline 
 1   &       --2.244    &     228.444     &        2.110   &      2.280     &    2.720 & \\
  2   &     5310.851   &    --5048.751   &        2.740   &      2.717   &      2.720 & \\
  3   &    10129.269  &     --9852.368    &       2.740  &       2.761   &      2.720 & \\ 
 4   &    14364.145  &    --14184.244     &      2.830  &       2.788   &      2.720 & \\
  5    &   18360.764  &    --18046.264   &        2.830   &      2.818   &      2.720 & \\
   6   &    21756.344   &   --21440.344   &        2.830  &      2.840   &      2.920 & \\
   7   &   24652.648   &   --24368.449    &       2.830   &      2.873   &      2.920 & \\
   8   &    27128.373  &   --26832.572   &        2.830   &      2.892  &       2.920 & \\
   9   &    29206.652   &   --28834.752    &       2.830   &      2.912   &      2.920 & \\
  10   &    30720.654   &   --30377.055   &        2.910  &       2.930  &       2.920 & \\
  11  &     31838.994  &    --31461.594    &       2.920   &      2.962   &      2.920 & \\
  12  &     32519.906   &   --32090.506     &      2.920   &        2.961  &       2.920 & \\
   13   &    32706.572   &   --32265.973   &        2.920   &        2.935   &      2.920 & \\
   14   &    32440.703  &    --31990.203    &       2.920   &        2.939  &      2.920 & \\
   15    &   31693.746   &   --31265.445     &      2.830  &         2.920   &      2.920 & \\
  16   &    30513.578   &   --30093.979    &       2.830    &       2.911  &       2.920 & \\
  17  &     28977.512   &   --28478.111       &    2.830    &       2.897  &       2.920 & \\
  18   &    26946.592   &   --26420.191    &       2.830   &        2.881  &       2.920 & \\
  19   &    24466.488   &   --23922.588    &       2.830   &        2.861   &      2.920 & \\
  20   &    21628.814   &   --20987.715   &        2.830    &      2.845   &      2.920 & \\
  21  &     18252.004  &    --17618.004    &       2.810   &        2.831   &      2.720 & \\
  22    &   14566.324   &   --13815.924   &        2.810    &       2.811   &      2.720 & \\
  23   &    10398.169  &     --9583.969   &         2.760   &        2.797    &     2.720 & \\
  24    &    5860.664   &    --4924.664     &       2.760     &      2.777  &       2.720 & \\
  25   &     1468.962    &     159.438      &       2.760   &        2.721   &      2.720 & \\ \hline \hline
 \end{tabular}  
 \caption{The columns describe  
point number, depth, {\it i.e.}, the distance below earth's crust (m), sea level height (m) 
(usually negative), and the densities from Crustal, from Shen-Rittzwoller, and from PEMC in
${\rm gm/cm^3}$.
}
\label{Table 2}
\end{center}
\end{table}
%
%
\begin{figure}
\includegraphics[width=.5\textwidth]{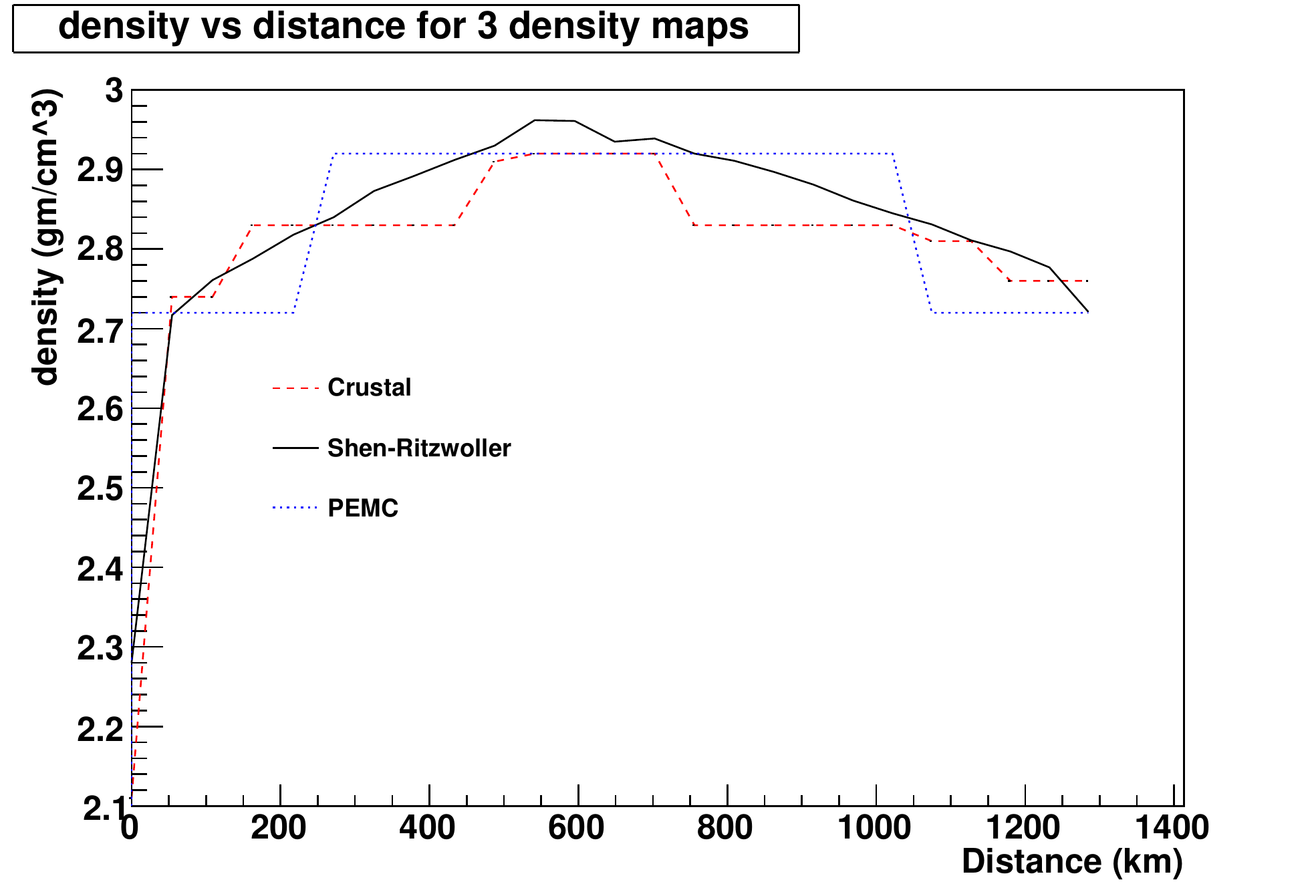}
\caption{ Densities vs distance.  The dashed line (red) is the CRUSTAL map, 
the solid line (black) is the Shen-Ritzwoller map,
and the dotted line (blue) is the old PEMC map.
}
\label{Figure 3}
\end{figure}

Although the actual situation is more complicated, we will look at uncertainties in the total
amount of matter passed through by the neutrinos ($\int\rho dx$) to get an indication of uncertainties.  There are two kinds
of uncertainties to be considered, statistical and systematic.  Statistical uncertainties are due
to random differences.
Sometimes the depths are near a boundary between two densities.
The boundaries are probably not completely flat and there is
some transition region.
In the crustal map there are six points within about 1.5 km of a depth boundary with an average 
change in density of about 4\%.
If we view this as a random walk then the standard deviation in the total amount of matter passed through
is 0.43\%.  Even if all twenty-five path segments had a 4\% uncertainty, the 
standard deviation in the total amount of matter passed through would be 0.8\%.  The statistical
uncertainties are quite small.

There are many more layers given for the  Shen-Ritzwoller map
and the differences from layer to layer are of the order of 
1\% (except for the last point, which has 15\% differences).  The statistical uncertainties are
again small.

The systematic uncertainties are those due to a systematic error in the density of the layers.
One approach is to compare the mean density for the three maps. The mean density for 
PEMC is  2.845 gm/cm$^3$ for  Crustal it is 2.817 gm/cm$^3$ and for  Shen-Ritzwoller 
it is 2.848 gm/cm$^3$. 
The PEMC map and the Shen-Ritzwoller map have essentially identical means while
the Crustal mean is approximately 1\% lower.  

Some early DUNE calculations used a mean density of 2.957 gm/cm$^3$ and a distance
of 1300 km \cite{Bass:2013vcg}.  This density is 4\% higher than the Shen-Ritzwoller mean density and
5\% higher than the crustal mean density.  In addition, the distance is 1\% longer than the
distance calculated here (1284.9 km), so the total amount of material through which
the beam passes is 5\% or 6\% higher
than the numbers here.

For the Shen-Ritzwoller map there is another way to estimate errors.  
They are still calculating detailed systematic errors, but they suggest that a reasonable
estimate of the error in density is to use the standard deviation in shear velocity ($v_s$) given in their Fig. 15 together with the
empirical relation between $v_s$ and $\rho$ obtained by T.M. Brocher \cite{Brocher},
\begin{align}
\rho = 1.227 +1.53v_s-0.837v_s^2+0.207v_s^3-0.01066v_s^4.
\end{align}
In their Fig. 15, the standard deviation in the magnitude of $v_s$ is of the order of 0.03 to 0.05
km/sec over the region of the DUNE beam.  The fractional errors in density 
obtained are fairly constant over the beam path.  
For 0.03, 0.05, and 0.07 km/sec errors in $v_s$, one obtains mean fractional errors 
in density of  0.5\%, 0.8\% and 1.2\%. 

\subsection{Electron Density Distribution In The Earth}
For a single kind of atom with atomic number $Z$ and given atomic weight, 
the number of atoms in one gm-atomic
weight is Avagadro's number ($N_{Av}$).  Let $\rho=$ the density of the
material in gm/cm$^3$. The number of electrons in one cubic centimeter $(N_e)$  is then
\begin{align}
N_e = Z\times N_{Av}\times \rho/{\rm atomic\ wgt}.
\end{align}
For a mix of materials the quantity needed is the mean value of $Z /{\rm atomic\ wgt}$.
Tables of the abundance in parts per million (ppm) of the various elements in the crust are given in
reference \cite{crustcomp}. In fact this reference lists three tables of 
abundances [15--17].  The
tables are in reasonable agreement for the main components, but
some of the minor elements differ by 20\% or more.
Table III gives the abundances for the most abundant 9 elements.
in ppm.  The further
elements are present only at the level of $<0.3$\%.  (Fe is the most abundant element
at lower depths, but not at the depths appropriate to this beam.)
\begin{table} [hp]
\begin{center}
\begin{tabular} {|  l  l | l  l   l | l  l  |} \hline \hline
Element & Z & \cite{abundanceone} & \cite{abundancetwo} & \cite{abundancethree} & Mean  & \\ \hline \hline  
 O &  8 &   460000. & 467100. & 461000. & 462700.    &  \\
 Si  &  14 &  270000. & 276900.  & 282000. & 276300.  &   \\        
  Al  &  13 &   82000. & 80700. & 82300.   &  81667.  &        \\  
  Fe  &  26 &  63000. & 50500. & 56300.   &    56600.  &       \\   
  Ca  &  20 &   50000. & 36500. & 41500.  &  42667.    &     \\      
  Na  &   11 &  23000. & 27500. & 23600.   &   24700.  &    \\     
  K    &   19 &  15000. & 25800. & 20900.   &  20567.   &    \\   
  Mg  &  12 &   29000. & 20800. & 23300.  &   24367.    &  \\      
 Ti    &  22 &  6600. & 6200. & 5600.         &   6133.    &   \\ \hline \hline 
\end{tabular}  
\caption{Abundances in ppm of the major elements in the Earth's crust.
}
\label{Table 3}
\end{center}
\end{table}

\begin{table} 
\begin{center}
\begin{tabular}   {| l  || l    l   | l    l    | l    l   | l     l   | l     l    l |} \hline \hline
El. & A & wgt & A & wgt & A & wgt & A & wgt & A & wgt &   \\ \hline \hline  
O & 16 & 15.995 & 17 & 16.999 & 18 &17.999 &  & &   &  &     \\     
Si & 28 & 27.977 & 29 & 28.976 & 30 & 29.974 &  &  &  & & \\             
Al & 27 & 26.982 &  &  &  &  &  & & &    &                  \\       
 Fe  & 54 & 53.940 & 56 & 55.935 & 57 & 56.935 & 58 & 57.933 &  &   &   \\          
 Ca   & 40 & 39.963 & 42 & 41.959 & 44 & 43.955 &  &  & & &        \\                 
  Na   & 23 & 22.990 &  &  &  &   & & & & &        \\                               
  K   & 39 & 38.96 & 41 & 40.962 &  &  &  & & & &        \\                       
  Mg   & 24 & 23.985 & 25 & 24.986 & 26 & 25.983 & &  &     &    &   \\                 
   Ti  & 46 & 45.953 & 47 & 46.952 & 48 &  47.948 & 49 &  48.948 & 50 &  49.945  &  \\   \hline \hline
\end{tabular} 
\caption{Isotopic numbers (A) and isotopic weights of stable isotopes of the major elements in the Earth's crust.
}
\label{Table 4}
\end{center}
\end{table}

\begin{table} 
\begin{center}
\begin{tabular}   {| l  || l    l   | l    l    | l    l   | l     l   | l     l    l |} \hline \hline
El. & A & abund & A & abund& A & abund & A & abund & A & abund &   \\ \hline \hline  
O & 16 & 99.757 & 17 & 0.038 & 18 &  0.205 & &  &  & &    \\     
Si & 28 & 92.230 & 29 & 4.683 & 30 & 0.0872 &  &  &  & & \\             
Al & 27 & 100. &  &  &  &  &  & & &    &                  \\       
 Fe  & 54 & 5.845 & 56 & 91.754 & 57 & 2.119 & 58 & 0.282 &  &   &   \\          
 Ca   & 40 & 96.941 & 42 & 0.647 & 44 & 2.086 &  &  & & &        \\                 
  Na   & 23 & 100. &  &  &  &   & & & & &        \\                               
  K   & 39 & 93.258 & 41 & 6.730 &  &  &  & & & &        \\                       
  Mg   & 24 & 78.99, & 25 & 10.0 & 26 & 11.01 & &  &     &    &   \\                 
   Ti  & 46 & 8.25 & 47 & 7.44 & 48 &  73.72 & 49 &  5.41 & 50 &  5.18  &  \\   \hline \hline
\end{tabular}  
\caption{Percentage isotopic abundances of stable isotopes of the major elements in the Earth's crust.
}
\label{Table 5}
\end{center}
\end{table}
 
 In addition the abundance of stable isotopes and atomic weights of these nine elements
are needed \cite{isotopeabund}.  The atomic weights  are given in Table IV and the
percentage fractional isotopic abundances in Table V.  Table VI gives Z$/$atomic weight and
Z$/$A averaged over the elements for each of the three abundance tables 
and for the mean, as well as the standard
deviation from the three tables.
\begin{table} 
\begin{center}
\begin{tabular}   {|   l  |  l |  l |  l |  l |  l |  l |} \hline \hline
                  & \cite{abundanceone} & \cite{abundancetwo} & \cite{abundancethree} & Mean & 
                  $\sigma$ \\ \hline
Z$/$wgt      &   0.4948 & 0.4950 & 0.4945 & 0.4949 & $1.013\times 10^{-4}$  \\ \hline
Z$/$A         &   0.4945 & 0.4947 & 0.49468 & 0.4946  & $1.030 \times 10^{-4}$   \\ \hline \hline 
\end{tabular}  
\caption{ Average Z$/$atomic weight, and Z$/$A, using the three different abundance tables.
The fourth column is  the result for the mean abundance from the three tables 
and the fifth column is the standard deviation
 of the three values.
}
\label{Table 6}
\end{center}
\end{table}

For the mean abundance, the number of electrons per cubic centimeter for $\rho = 1$ is 2.9805$\times 10^{23}$.
The fact that $Z/A$ is so near to 1/2 is not surprising.   The most abundant elements, 
oxygen (O) and silicon (SI), comprising about 75\% of the total have isotopic abundances overwhelmingly favoring
1/2.
\section{Neutrino Oscillation Probabilities at Sanford Laboratory}
For the present analysis, the density results using the new Shen-Ritzwoller  map are used
with one small modification.  It was more convenient to
have the neutrino beam distances between points constant for the density vs distance map.  Here
the average distance between points was used.  The maximum distance change was about
6 km.  That occurred at the center of the path, where the density changes from point
to point are small.  

Neutrino oscillations are calculated for a variable density path using the
computer program of J. Kopp \cite{kopp}.
Results are presented for this variable density map, for a constant density of 2.848 gm/cm$^3$, 
which is the mean density for this variable density map, and for the density of 2.957 gm/cm$^3$.  The distance between FNAL and 
the Sanford Laboratory was calculated in Section II to be 1284.9 km.  The DUNE
calculations which used 2.957 gm/cm$^3$ used 1300 km as a distance.  For the present 
comparison a distance of 1284.9 km was used for this density as well.  

%

\subsection{Plots of Oscillation Probabilities for the Variable Density Option}
Figures 4--7 show plots of oscillation probabilities at Sanford Laboratory  for $\nu$ and $\bar{\nu}$ oscillations separately, for both the  CP violation parameter $\delta_{CP}=0$ and $\delta_{CP} = 3\pi/2$. The differences between the three density options, 
for $\nu$ and $\bar{\nu}$, for $\delta_{CP} = 0$ 
and $\delta_{CP} = 3\pi$/2  have been calculated, 
As an example, the differences between the variable density
option and the fixed 2.848 gm/cm$^3$ density option for $\nu$ and $\bar{\nu}$
with $\delta_{CP}=0$ are shown in Figs. 8 and 9.  Note the difference of probability  scales
between Figs. 4 to 7 and Figs. 8-9.
\begin{figure} [htbp]
   \includegraphics[width=0.45\textwidth]{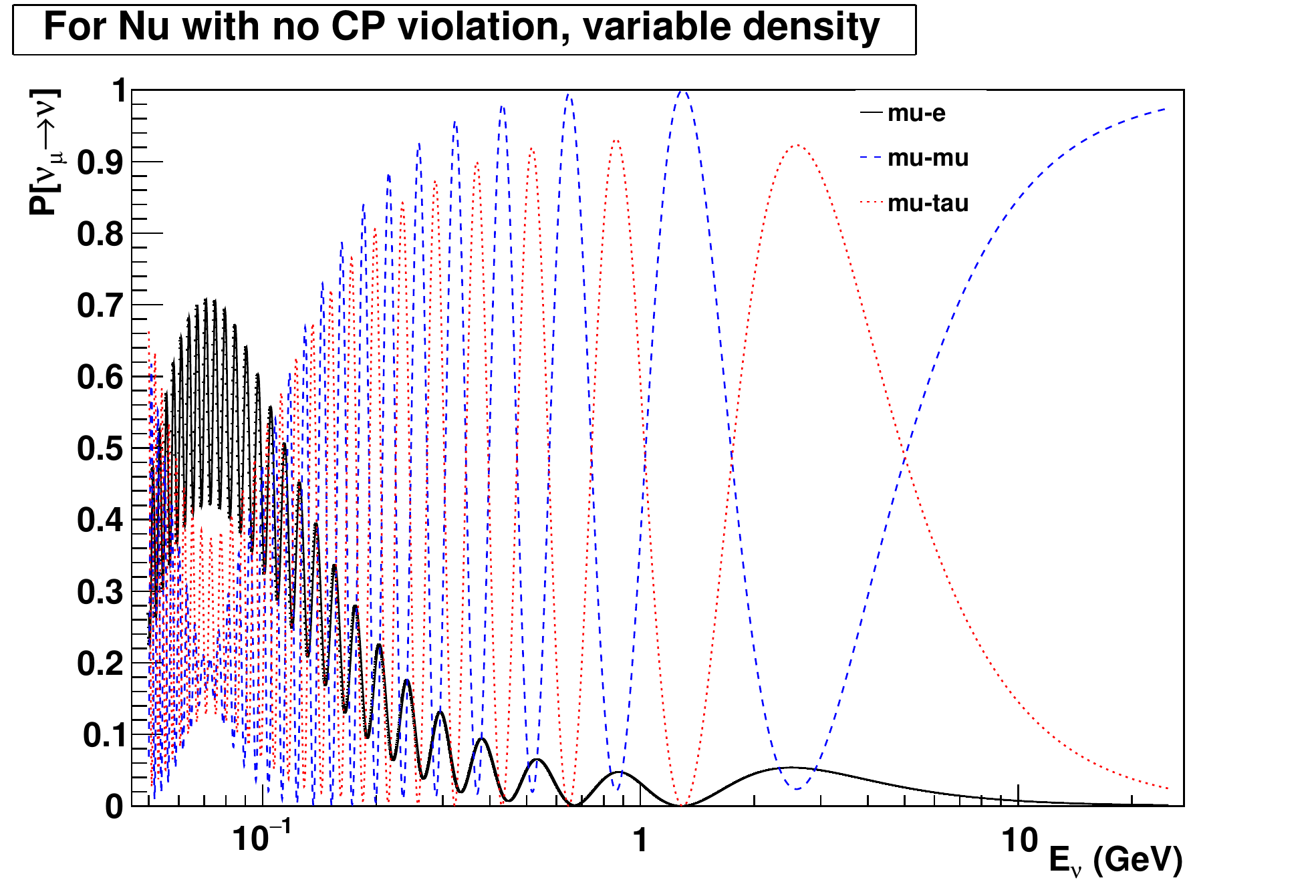}
  \caption{Pr($\nu$) oscillations with $\delta_{CP} = 0$  using the variable density path.}
   \label{ 4} 
\end{figure}
\begin{figure}
   \includegraphics[width=0.45\textwidth]{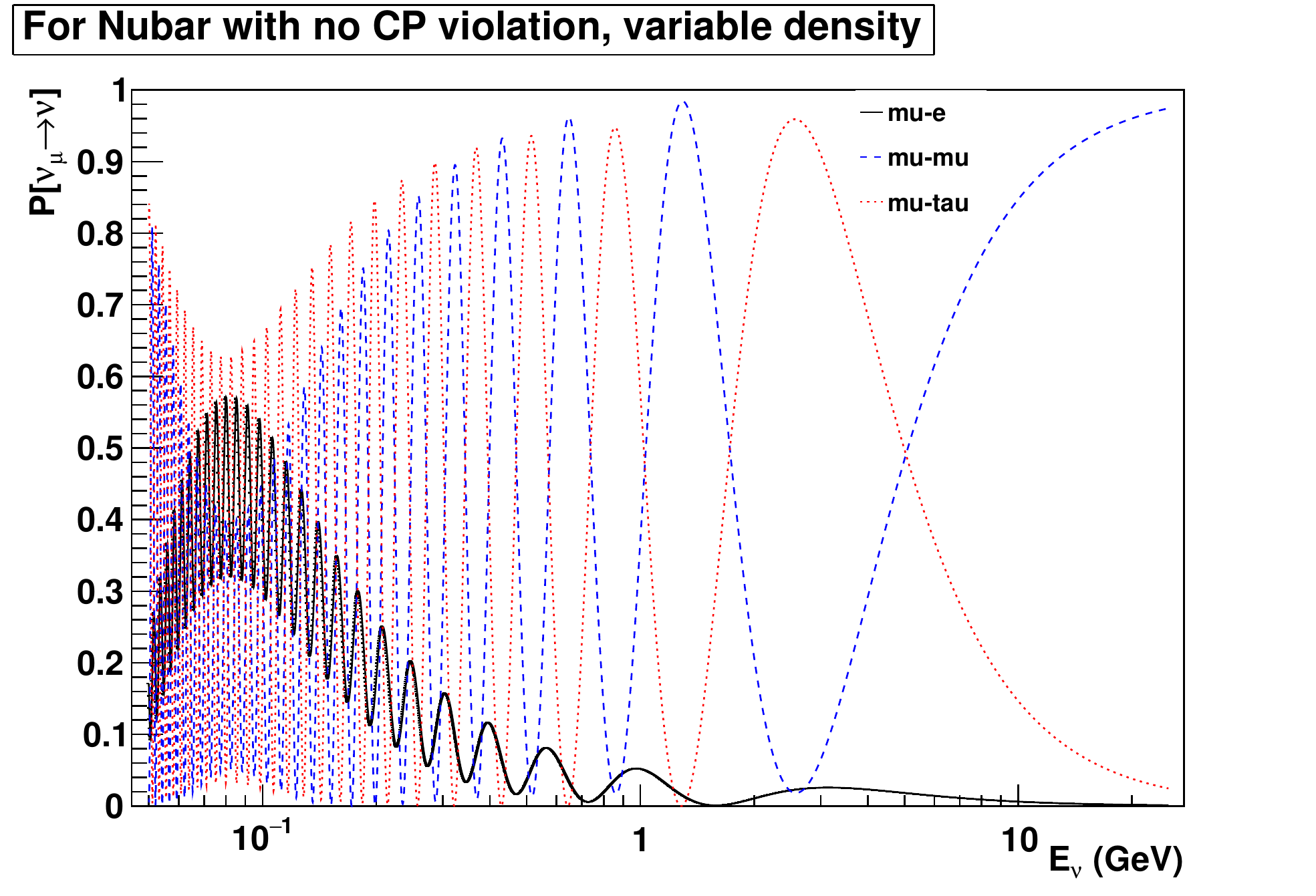}
   \caption{Pr($\bar{\nu}$) oscillations with $\delta_{CP} = 0$ using the variable density path.}
   \label{ 5}
\end{figure}
\begin{figure} [htbp]
   \includegraphics[width=.45\textwidth]{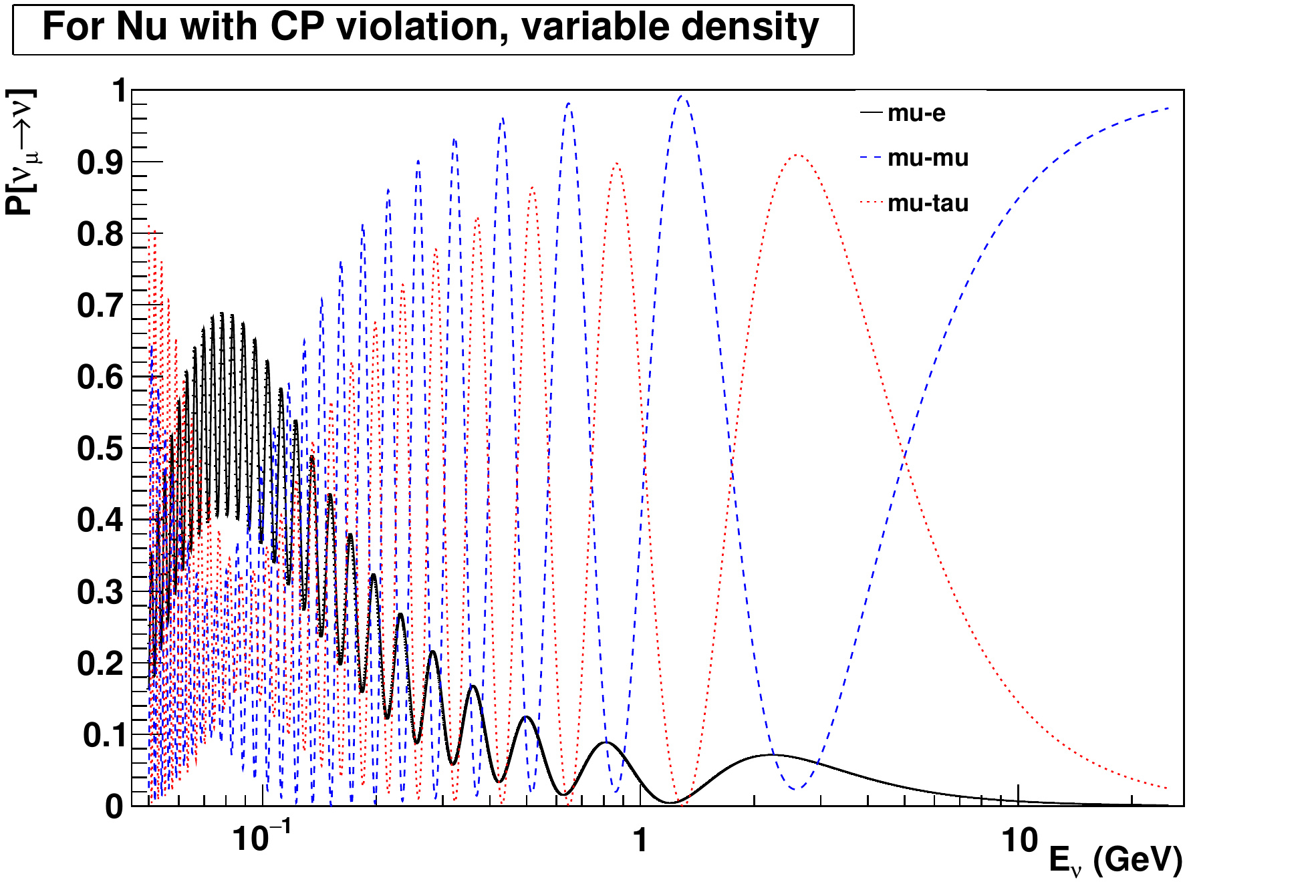}
  \caption{Pr($\nu$)  oscillations with $\delta_{CP} = 3\pi$/2 using the variable density path.}
   \label{ 6} 
\end{figure}
 \begin{figure} [htbp]%
   \includegraphics[width=.45\textwidth]{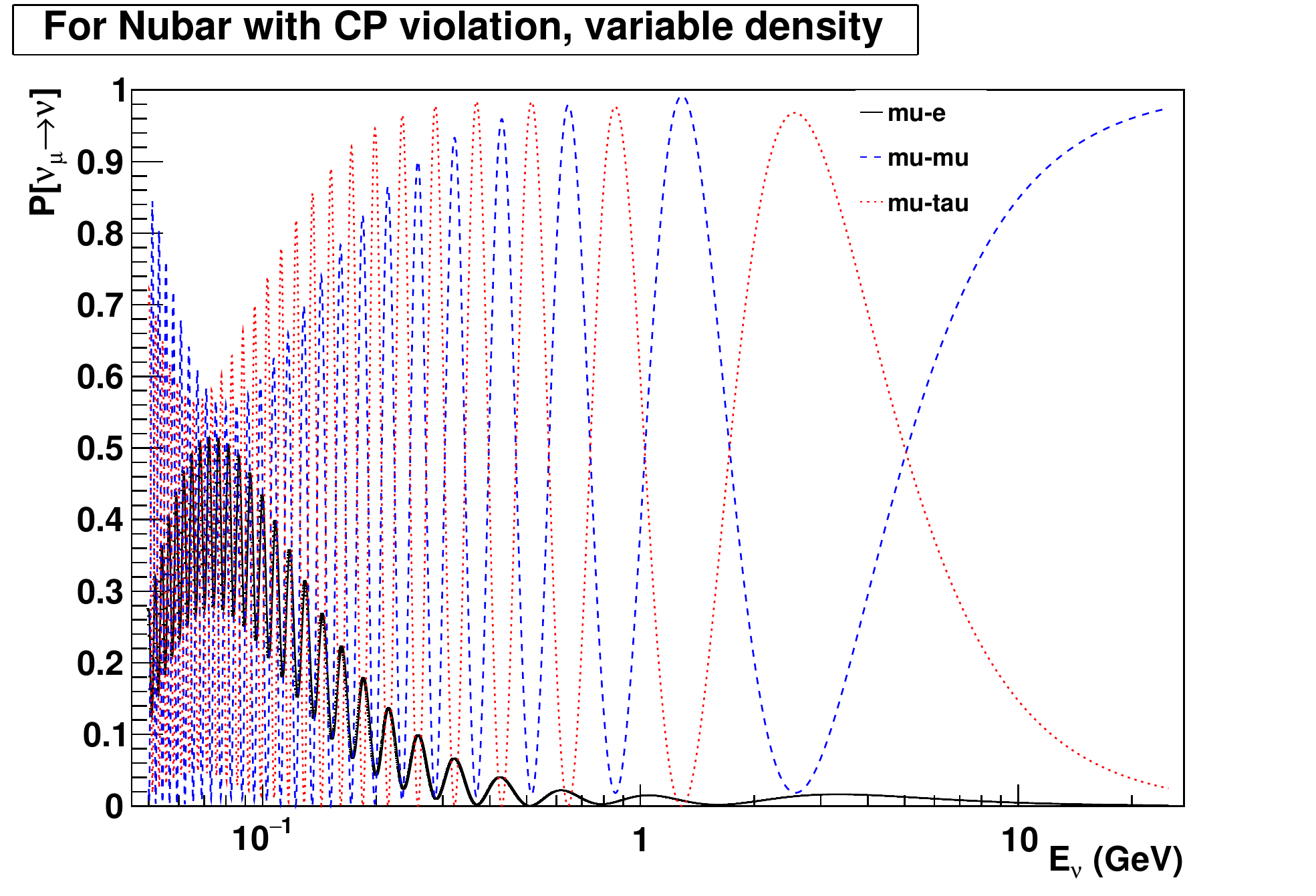}
   \caption{Pr($\bar{\nu}$) oscillations with $\delta_{CP} = 3\pi$/2  using the variable density path.}
   \label{ 7}
\end{figure}
\begin{figure} [htbp]
   \includegraphics[width=.45\textwidth]{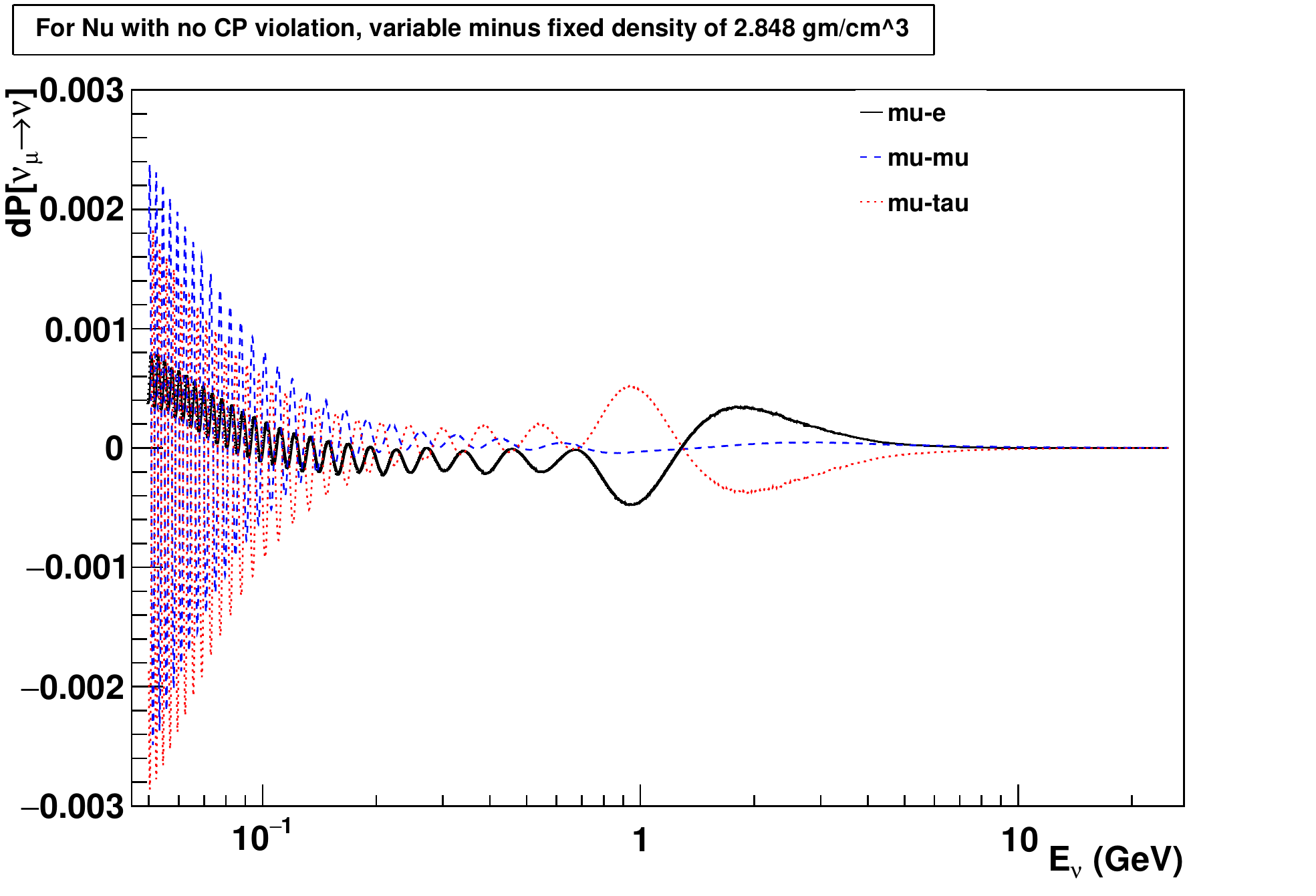}
  \caption{Pr($\nu$)  oscillations with $\delta_{CP} = 0$; variable density minus fixed density of 2.848 gm/cm$^3$.}
   \label{ 8} 
\end{figure}
 \begin{figure} [htbp]%
   \includegraphics[width=.45\textwidth]{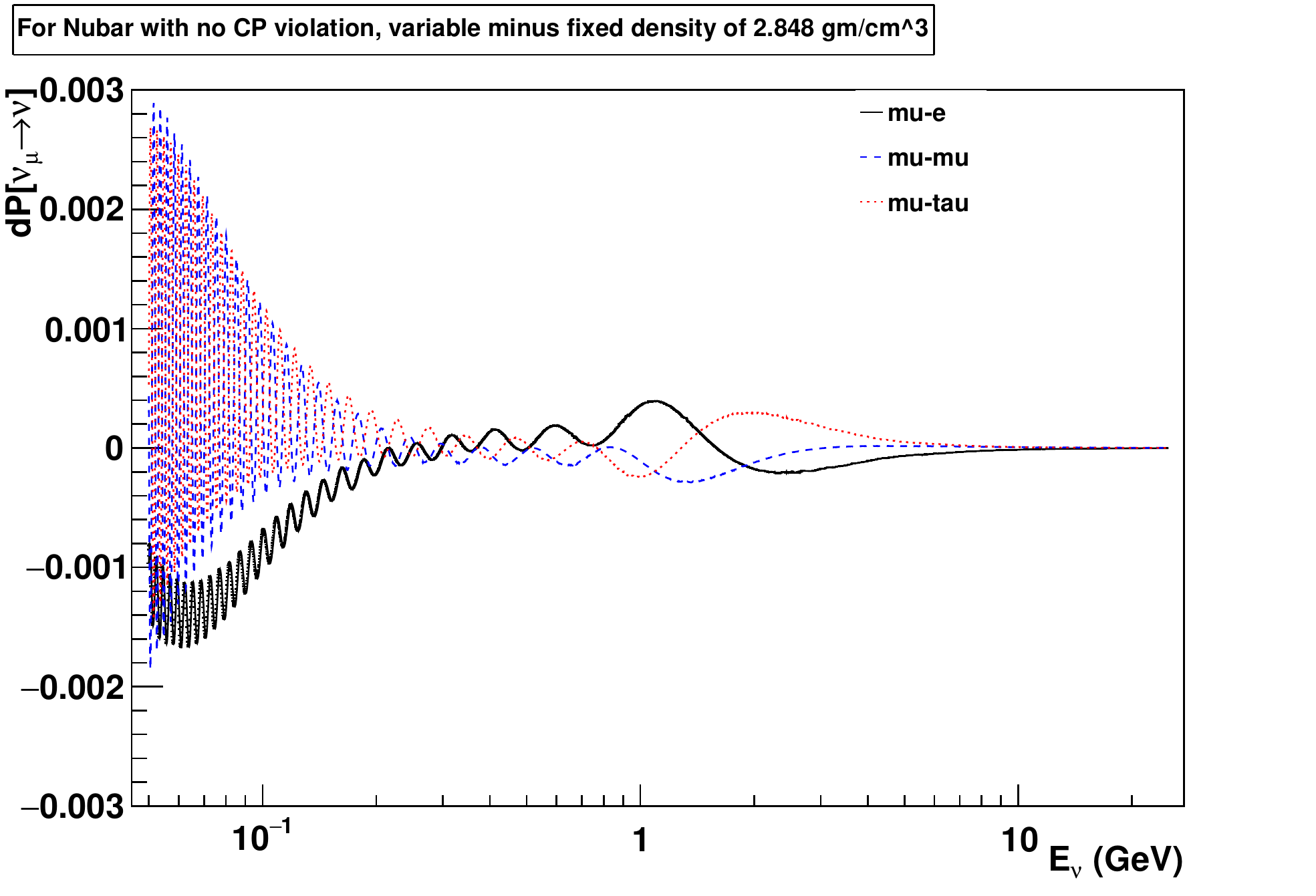}
   \caption{Pr($\bar{\nu}$) oscillations with $\delta_{CP} = 0$; variable density minus fixed density of 2.848 gm/cm$^3$.
   }
   \label{ 9}
 \end{figure}

\subsection{Discussion Of Results}

Selected Pr($\nu_e$) and Pr($\bar{\nu}_e)$ oscillation peaks near 0.1 GeV, 0.5 GeV, and 0.8--1.3 GeV were compared
for the three density assumptions, and for both $\delta_{CP}=0$ and $\delta_{CP} = 3\pi/2$, 
a total of 36 comparisons. 
The experimental flux is negligible in the region around 0.1 GeV and it is included only
to give a sequence of energies encompassing much of the the experimentally interesting  region.
In practice energy bands will have to be selected.
However,
an average will likely reduce the differences and be very
dependent on the kind and range of the average. For the present purpose,
this is avoided.

In all of the 36 density comparisons for $\nu_e$ and $\bar{\nu}_e$ the
locations of the peaks in energy were identical within 0.3\% for the different density 
assumptions.  In one comparison  the difference in peak size was 1.4\%.  In all
other comparisons the size difference was $< 1$\%.  For $\nu_\mu$ and $\bar{\nu}_\mu$
the maximum energy location difference was  $<0.3$\% and
the peak size differences were $<0.65$\%.  These are quite small differences.

Nonetheless some comparisons were made for two quantities that might be used to
look at matter effects and 
CP violation to see if any subtle differences might appear.  The first quantity was:
\begin{align}
\Delta_1(E) = {\rm Pr}(\nu_e) -{\rm Pr}(\bar{\nu}_e).
\end{align}
$E$ is the energy at which the comparison is made.
Since $\nu$ and $\bar{\nu}$ behave differently under interactions with matter, $\Delta_1$
serves to emphasize the matter interactions. 

The second quantity examined was:
\begin{align}
\Delta_2(E)=({\rm Pr}( \nu_e)-{\rm Pr}(\bar{\nu}_e)){\rm\ for\ }\delta_{CP} = 0 \nonumber \\ 
- ({\rm\ Pr}( \nu_e)-{\rm Pr}(\bar{\nu}_e))
{\rm\  for 
\ }\delta_{CP} = 3\pi/2.
\end{align}
This is an important quantity to use to look at CP violation.

$\Delta_1(E)$ and $\Delta_2(E)$
were examined for each of the three density assumptions, and $\Delta_1(E)$ was examined both 
for $\delta_{CP}=0$ and for $\delta_{CP} = 3\pi/2$ .


In Table VII, for the variable density assumption, 
three energies corresponding to probability maxima for $\Delta_1$ and $\Delta_2$ 
are shown along with their  maximum values.  

For the following tables, ``v" refers to the variable density assumption, ``s" refers to a fixed density of
$\rho=2.848$ gm/cm$^3$, and ``d" refers to a 
fixed density of $\rho=2.957$ gm/cm$^3$.  
``$v-s$" means variable density minus fixed density 2.848 gm/cm$^3$, 
 ``$v-d$" means variable density minus fixed density 2.957 gm/cm$^3$, and ``$d-s$" means
 fixed density 2.957 gm/cm$^3$ minus fixed density 2.848 gm/cm$^3$. 
 For comparisons involving the variable density, the energies correspond
 to the three energy values in Table VII.
For the comparisons of ``$d-s$" the values for the peak energies for ``d" nearest to those in Table VII were chosen.

Table VIII  examines the differences between
the $\Delta_1(E_{Peak1})$ values for the different density assumptions, where
$E_{peak1}$ is the energy of the 
maximum $\Delta_1$ for the first density assumption.
$\delta(\Delta_1(E_{peak1}))$ is the difference of $\Delta_1$  found in the
two density assumptions. The percentages of the ratio 
 $\delta(\Delta_1(E_{peak1}))/\Delta_1(E_{peak1})$ 
 are shown for each of the three energies.

In Tables IX and X,   $E_{max1}$ is the nearest energy to $E_{peak1}$ for which
$|\delta(\Delta_1(E_{max1}))|$ is at a local maximum. 
The percentage differences of $E_{max1}$ from $E_{peak1}$ 
and of the ratio $\delta(\Delta_1(E_{max1}))/\Delta_1(E_{max1})$ 
are shown.  Table IX 
shows these quantities if
$\delta_{CP}=0$ and Table X shows these quantities if $\delta_{CP}=3\pi/2$.
The $\Delta_1$ differences are sometimes appreciable, although 
the values of $\Delta_1$ often are small.

Table XI  examines the differences between
the $\Delta_2(E_{Peak2})$ values for the different density assumptions.
$E_{peak2}$ is the energy of the 
 maximum $\Delta_2$ for the first density assumption.
  $\delta(\Delta_2(E_{peak2}))$ is the difference of $\Delta_2$  found in the
 two density assumptions. The percentages of the ratio 
 $\delta(\Delta_2(E_{peak2}))/\Delta_2(E_{peak2})$ 
 are shown for each of the three energies.

In Table XII,  $E_{max2}$ is the nearest energy to $E_{peak2}$ for which
$|\delta(\Delta_2(E_{max2}))|$ is at a local maximum. 
The percentage differences of $E_{max2}$ from $E_{peak2}$ 
and of the ratio $\delta(\Delta_2(E_{max1}))/\Delta_2(E_{max1})$ 
are shown. 

For $\Delta_2$, the difference
between using the variable density and the mean of the variable density, 2.848 gm/cm$^3$ is
small, of the order of 0.2\%, except for the one anomalous value.  That value occurs because
the largest value of $\delta(\Delta_2)$ is at a point where the new value of $\Delta_2$ is almost zero.  
In general the percent errors for $\Delta_2$ are less than those for $\Delta_1$.  
Some of the differences between the various density
assumptions cancel for $\Delta_2$.  It is worth noting that, even if a constant density is used,
a beam length of 1284.9 km should be used rather than 1300 km.

There may be other tests and energies which would show larger differences.  The
Kopp variable density routine (with some small modifications which were made to look
at a density vs distance graph), is reasonably easy to use and is very fast.
The 12 basic output files used for this paper (3 density choices, with $\delta_{CP}=0$ and $\delta_{CP} = 3\pi/2$,
and $\nu$ and $\bar{\nu}$) can be downloaded from my homepage \cite{roehome}.  

 \begin{table} 
 \vspace{5mm}
\begin{center}

\begin{tabular}      { | l  | l   | l   | l   | l   | l   | l  | l  l |}         \hline \hline 
  $\Delta$ & $\delta_{\rm CP}$ & E  & $\Delta$ & E & $\Delta$ & E & $\Delta$ & \\ \hline\hline
$\Delta_1$ & 0. & 0.096 & 0.14 & 0.42 & $-$.0521 & 1.12 & $-$0.028 & \\ \hline
$\Delta_1$ & 1.5$\pi$ & 0.096 & 0.41 & 0.37 & 0.16 & 0.811 & 0.086  & \\ \hline
$\Delta_2$ &     &  0.096 & -0.27 & 0.37 & $-$0.15 & 0.827 & $-$0.069 & \\ \hline\hline
 \end{tabular}  
 \caption{Results for the variable density option for $\Delta_1 = {\rm Pr}(\nu) -{\rm Pr}(\bar{\nu})$ and 
 $\Delta_2 = \Delta_1(\delta_{CP}=0) - \Delta_1(\delta_{CP} = 3\pi/2$). The columns
 labelled $\Delta$ are $\Delta_1$ or $\Delta_2$ as designated in column 1.  
 E is the energy of the chosen maximum $\Delta$
 in GeV.
}
\label{Table 7}
\end{center}
\end{table}
  \begin{table} 
\begin{center}
\begin{tabular}          {|  l |  l  |  l  |  l |   l  l | }     \hline \hline 
$\Delta$ var & $\delta_{CP}$ & $\delta(\Delta_1)$ & $\delta(\Delta_1)$ & $\delta(\Delta_1)$ &  \\ \hline\hline
$v-s$ & 0. & 0.92  & 0.54 & 2.47 & \\ \hline
$v-d$ & 0. & $-$2.6 & $-$2.8 & 0.68 &  \\ \hline
$d-s$ & 0. & 3.5 & 3.4 & $-$3.15 & \\ \hline
$v-s$ & $3\pi/2$ & 0.27 & 0.18 & 0.57 & \\ \hline
$v-d$ & $3\pi/2$ & 0.21 & 0.007 & 1.0 & \\ \hline
$d-s$ & $3\pi/2$ & 0.48 &$-$0.18 & 0.47  & \\ \hline\hline
 \end{tabular}  
 \caption{$\Delta_1(E_{peak1}) = {\rm Pr}(\nu) -{\rm Pr}(\bar{\nu})$. $E_{peak1}$ is the energy of the 
 maximum $\Delta_1$ for the first density assumption.
  $\delta(\Delta_1(E_{peak1}))$ is the difference of $\Delta_1$  found in the
 two density assumptions. The percentages of the ratio 
 $\delta(\Delta_1(E_{peak1}))/\Delta_1(E_{peak1})$ 
 are shown for each of the three energies.
 }
\label{Table 8}
\end{center}
\end{table}
  \begin{table} 
\begin{center}
\begin{tabular}      {  | l   | l   | l   | l   | l   | l  | l  l |}         \hline \hline 
$\Delta$ var & dE & $\delta(\Delta_1)$ & dE & $\delta(\Delta_1)$ & dE & 
 $\delta(\Delta_1)$ &  \\ \hline\hline
$v-s$ & 0 & 0.92 & $-$4.9 & $-$0.91 & -7.8 & 3.5 & \\ \hline
$v-d$ & 0 & $-$2.6 & 0.47 & $-$2.8 & 6.7 &0.98 &  \\ \hline
$d-s$ & 0 & 0.92 & 0 & 3.4 & $-$0.97 & 3.16 &  \\ \hline\hline
 \end{tabular}  
 \caption{$E_{peak1}$ is the energy of the maximum $\Delta_1$ for the
first density assumption and $E_{max1}$ is the nearest energy to $E_{peak1}$ for which
$|\delta(\Delta_1(E_{max1}))|$ is at a local maximum. 
The percentage differences of $E_{max1}$ from $E_{peak1}$ 
and of the ratio $\delta(\Delta_1(E_{max1}))/\Delta_1(E_{max1})$ 
are shown for each of the three energies.
 $ \delta_{CP}=0$ is assumed for  this table.
}
\label{Table 9}
\end{center}
\end{table}
  \begin{table} 
\begin{center}
\begin{tabular}      {  | l   | l   | l   | l   | l   | l  | l  l |}         \hline \hline 
$\Delta$ var & dE & $\delta(\Delta_1)$ & dE & $\delta(\Delta_1)$ & dE & 
 $\delta(\Delta_1)$ &  \\ \hline\hline
$v-s$ & $-$0.31 & 4.0 & 3.3 & 0.25 & 15.2 & 1.5 & \\ \hline
$v-d$ & $-$3.1 &10.1 & 8.4 & 1.2 & $-$9.0 &1.3 & \\ \hline
$d-s$ & $-$3.1 & 16.6 & 7.4 & 1.3 & $-$9.0 & 1.1 & \\ \hline\hline
 \end{tabular}  
 \caption{ $E_{peak1}$ is the energy of the maximum $\Delta_1$ for the
first density assumption and $E_{max1}$ is the nearest energy to $E_{peak1}$ for which
$|\delta(\Delta_1(E_{max1}))|$ is at a local maximum. 
The percentage differences of $E_{max1}$ from $E_{peak1}$ 
and of the ratio $\delta(\Delta_1(E_{max1}))/\Delta_1(E_{max1})$ 
are shown for each of the three energies.
 $ \delta_{CP}=3\pi/2$ is assumed for  this table.
 }
\label{Table 10}
\end{center}
\end{table}
\begin{table} 
\begin{center}
\begin{tabular}          {|   l  |  l  |  l |   l  l |}     \hline \hline 
$\Delta$ var & $\delta(\Delta_2)$ & $\delta(\Delta_2)$ & $\delta(\Delta_2)$ & \\ \hline \hline
$v-s$ & $-$0.05 & $-$0.62 & $-$0.31 & \\ \hline
$v-d$ & 0.95 & 0.40 & 0.20 & \\ \hline
$d-s$ & $-$1.0 & $-$0.53 & $-$0.37  & \\ \hline\hline
 \end{tabular}  
 \caption{ 
 $\Delta_2(E_{peak2}) = \Delta_1(\delta_{CP}=0) - \Delta_1(\delta_{CP} = 3\pi/2$). $E_{peak2}$ is the energy of the 
 maximum $\Delta_2$ for the first density assumption.
  $\delta(\Delta_2(E_{peak2}))$ is the difference of $\Delta_2$  found in the
 two density assumptions. The percentages of the ratio 
 $\delta(\Delta_2(E_{peak2}))/\Delta_2(E_{peak2})$ 
 are shown for each of the three energies.
 }
\label{Table 11}
\end{center}
\end{table}

  \begin{table} 
 
\begin{center}
\begin{tabular}      {  | l   | l   | l   | l   | l   | l  | l  l |}         \hline \hline 
$\Delta$ var & dE & $\delta(\Delta_2)$ & dE & $\delta(\Delta_2)$ & dE & 
 $\delta(\Delta_2)$ &  \\ \hline\hline
$v-s$ & 4.6 & $-794.$ & $-0.62$ & $-0.13$ & $-4.0$ & $-0.20$ & \\ \hline\hline
$v-d$ & 6.3 & 1.1 & 1.6 & 0.48 & 9.7 & 0.53  & \\ \hline
$d-s$ & 6.3 & 1.2 & 1.4 & $-$0.59 & 6.7 & $-$0.56 & \\ \hline\hline
 \end{tabular}
\caption{$E_{peak2}$ is the energy of the maximum $\Delta_2$ for the
first density assumption and $E_{max2}$ is the nearest energy to $E_{peak2}$ for which
$|\delta(\Delta_2(E_{max2}))|$ is at a local maximum. 
The percentage differences of $E_{max2}$ from $E_{peak2}$ 
and of the ratio $\delta(\Delta_2(E_{max2}))/\Delta_2(E_{max2})$ 
are shown for each of the three energies.
}
\label{12}
\end{center}
\end{table}


\begin{acknowledgements}
I wish to acknowledge the considerable help of Professor Henry Pollack of the Earth and 
Environmental Sciences Department, University of Michigan,
in providing considerable expertise to help me understand at least some elementary basics
of the field.  I wish to thank Professor Joshua Spitz, Department of Physics, University of Michigan 
for help in obtaining the Shen-Ritzwoller density tables, for introducing me to the Kopp program 
and for help in using it.

\end{acknowledgements}
{}

\end{document}